\documentclass[12pt,preprint]{aastex}

\shorttitle{Hot Fast Flow above a Flare Arcade}
\shortauthors{Imada et al.}

\begin{document}

\title{EVIDENCE FOR HOT FAST FLOW ABOVE A SOLAR FLARE ARCADE}

\author{S. \textsc{Imada},\altaffilmark{1, 2} 
K.  \textsc{Aoki},\altaffilmark{2,3} 
H.  \textsc{Hara},\altaffilmark{2,3} 
T.  \textsc{Watanabe},\altaffilmark{2}
L. K. \textsc{Harra},\altaffilmark{4}
T. \textsc{Shimizu} \altaffilmark{5}}
  
\altaffiltext{1}{ Solar-Terrestrial Environment Laboratory (STEL), Nagoya University, Furo-cho, Chikusa-ku, Nagoya 464-8601, Japan}
\altaffiltext{2}{ National Astronomical Observatory of Japan,  2--21--1 Osawa, Mitaka-shi, Tokyo 181--8588, Japan}
\altaffiltext{3}{ Department of Astronomy, University of Tokyo, Hongo Bunkyo-ku, Tokyo 113-0033, Japan}
\altaffiltext{4}{ UCL-Mullard Space Science Laboratory, Holmbury St. Mary, Dorking, Surrey RH5 6NT, UK}
\altaffiltext{5}{Institute of Space and Astronautical Science, Japan Aerospace Exploration Agency, 3--1--1 Yoshinodai, Sagamihara-shi, Kanagawa 229--8510, Japan}

\begin{abstract}
Solar flares are one of the main forces behind space weather events. However the mechanism that drives such energetic phenomena is not fully understood. The standard eruptive flare model predicts that magnetic reconnection occurs high in the corona where hot fast flows are created. Some imaging or spectroscopic observations have indicated the presence of these hot fast flows but there have been no spectroscopic scanning observation to date to measure the two-dimensional structure quantitatively. We analyzed a flare that occurred on the west solar limb on 27 January 2012 observed by the Hinode EUV Imaging Spectrometer (EIS) and found that the hot ($\sim$30MK) fast ($>$500 km s$^{-1}$) component was located above the flare loop. This is consistent with magnetic reconnection taking place above the flare loop.
\end{abstract}


\section{INTRODUCTION}

A solar flare is a sudden brightening observed in almost all wavelengths. The energy released by a flare is so huge that the total amount of energy often reaches 10$^{32}$ erg within an hour. Solar flares are sometimes associated with coronal mass ejections (CMEs) which can trigger geomagnetic storms. These sudden huge energy releases can also be observed on other magnetized astronomical objects. Therefore, over the last several decades, considerable effort has been devoted towards understanding how to convert magnetic energy to plasma energy during solar flares. The so-called 'standard model' of eruptive flares which is based on magnetic reconnection have been proposed (the CSHKP model \cite{car,stu,hir,kop}), and some of the predicted characteristics have been verified by observations (e.g., cusp-like structure in soft X-ray images: \cite{tsu}, hard X-ray source above the flare loop: \cite{mas}, chromospheric evaporation: \cite{ter,ima2}, reconnection inflows: \cite{yok}, reconnection outflows (off limb: \cite{inn,liu}, on disc: \cite{har}), plasmoid ejection: \cite{ohy}, and CMEs: \cite{sve, ima, ima4}).   

 One of the typical characteristics predicted by CSHKP model is that magnetic reconnection occurs above the flare arcades. Therefore, hot (a few 10 MK) and fast ($\sim$1000 km s$^{-1}$) plasma flows should be observed above the flare arcades. The magnetic reconnection region is sandwiched between two slow-mode shocks and hence should appear as a narrow structure (like a spear). 
However, other physical mechanisms such as a blast wave \citep[][]{inn2,tot} can explain the  presence of hot fast flows above the flare arcade, so it is crucial to observe the hot fast flow structure quantitatively in order to understand the solar flare process.
So far, many observations at the solar limb have been made to confirm the presence of hot fast flows above the flare arcade. Observations at the limb have the advantage that the height information can be clearly determined. For example, supra-arcade downflows (SADs) are observed by coronal imagers which are interpreted as the reconnection outflows predicted from the flare model. X-ray dark voids and sometimes bright features are observed to move sunward (downward) from the high corona with apparent velocity of a few 100 km s$^{-1}$, especially during the late phase of flares \citep[][]{mck,mck2,sav}. However, equivalent spectroscopic observations are very rare because of the slow spatial scanning and small field of view. Only a few spectroscopic limb observations have reported the presence of hot fast flows above the flare arcade \citep[][]{inn,wan}. These studies are from a single slit position above the flare arcades, and hence two-dimensional spatial information of the hot fast flows was not available. In this letter, we show the first spectral scanning observation of hot fast flow above a flare seen on the solar limb.
 
\section{OBSERVATION AND DATA ANALYSIS}

On 27 January 2012, Hinode \citep[][]{kos} observed a large solar flare (GOES X1.7, peak time 18:37) at the northwest solar limb (30$^\circ$ N, 90$^\circ$ W). The EUV Imaging Spectrometer (EIS) aboard Hinode is a high spectral/spatial resolution spectrometer aimed at studying dynamic phenomena in the corona \citep{cul}. The Hinode EIS observed the flare with a slit scanning mode with a 2" wide slit and exposure duration of 5 s at each scanning point.
The observing period is from 18:16 to 18:29 UT, which is during the rise phase of the flare. 
This is the first raster scanning observation ($\sim 10$ minutes for one raster scan), before the fast sparse raster scanning begins ($\sim 5$ minutes cadence).
To process the EIS data we used the software provided by the EIS team (eis\_prep), which corrects for the flat field, dark current, cosmic rays, and hot pixels.
 During the flare, EIS successfully obtained EUV images and line-of-sight (LOS) velocities (Doppler shift) in several emission lines, including the \ion{Fe}{24} line at 192\AA (window width is $\sim 0.535$ \AA), with $\sim$1500 km spatial resolution.
By using the hot \ion{Fe}{24} emission line \citep[see CHIANTI,][]{der,lan}, we can reveal the fine-scale structure and dynamics of the high-temperature ($\sim10^7$K) plasma in the flaring region. 
We also have the other flare lines (e.g., \ion{Fe}{23} (263.77 \AA), \ion{Fe}{24} (255.11 \AA)) and cooler lines (e.g., \ion{Fe}{11} (188.23 \AA), \ion{Fe}{14} (274.20 \AA)) for this observation. 
These lines are useful to check the blending effect in the \ion{Fe}{24} line at 192 \AA.
The \ion{Fe}{24} line at 192 \AA~ is blended with \ion{Fe}{11} and \ion{Fe}{14}. 
By comparing the \ion{Fe}{24} at 192 \AA~ result and the other flare/cooler lines results, we can distinguish the flare plasma information from the blended emission.
Simultaneously, the Solar Dynamics Observatory (SDO)/ Atmospheric Imaging Assembly (AIA) instrument acquired full-Sun images with a spatial resolution of ~1000 km. We use the AIA spectral channels at 94, 13, and 193 \AA~to analyze the high temporal evolution of the flaring plasma ($\sim$24 s). Each of the AIA channels has different temperature coverage. The AIA 94, 131, and 193-\AA~channels represent 10$^{6.8}$, 10$^{7.0}$, 10$^{7.3}$ K plasma respectively in the case of flares as discussed by \cite[][]{boe}. 

The AIA images taken in the different channels show the behavior of the flare loop structure at different temperature (Figure 1). The development of the flaring loops can be clearly seen by the AIA movie (movie S1). To clarify whether the hot (a few 10 MK) Alfv\'enic flows ($\sim$1000 km s$^{-1}$) are located above the flaring loops or not, as predicted by CSHKP model, we analyze the AIA images. The downward (sunward) moving features above the flare arcades in the highest temperature range (193-\AA~channel) are clearly observed (movie S1). Even in the 131-\AA~channel we can also see some downward moving features, but in the 94-\AA~channel, as the temperature decreases, there are no clear signature of flows. This can be interpreted as evidence for the hot downward moving loops. We can estimate the temperature from the intensity ratio between 193 and 131-\AA~channel, and we obtain that the downward moving loops are roughly $\sim$30 MK. To estimate the apparent velocity of downward moving loops, we create a time-distance plot along the white line in Figure 2a. The time-distance plot (Figure 2b) shows patterns in the flow with transverse velocity with a typical value of 350 km s$^{-1}$. In Figure 2b, we can also see the development of the bright flare loops associated with the downward moving features. 

 EIS \ion{Fe}{24} (192 \AA) images are shown in Figure 3. 
 The line-center image of \ion{Fe}{24} (Figure 3b; stationary in LOS direction, $192.03 \pm 0.156$ \AA) shows the flare loops which are similar to those observed in AIA 193-\AA~channel. 
 This indicates that the AIA 193-\AA~images are showing structures filled with $>$10MK plasma. 
 The blue-wing $191.8 \pm 0.0223$ \AA~image (Figure 3a), which corresponds to plasma flowing toward Hinode with a velocity of 400 km s$^{-1}$, shows a totally different shape from the line-center image. The red-wing $192.6 \pm 0.0223$ \AA~image is also shown in Figure 3c. We have compared Figure 3a-c with the figure made by the other flare lines or cooler lines in the same way, and confirm that the Figure 3a-c are certainly the results from \ion{Fe}{24}.
In the line center image, we see a cross-shaped fringe pattern which is due to diffraction.
There is  no diffraction pattern in the blue/red-wing image, because the bright flare arcade do not have strong blue/red-wing component.
We think this is one of the reasons why we can see a lot of structure in the blue/red-wing image.
 In the northeastern part of flare, where AIA observations show the hot and fast apparent flow, we can see the diffuse structures in the blue-wing image (Figure 3a), which cannot be seen in the line-center image (Figure 3b). This diffuse structure, seen only in the blue-wing image, is indicating a response to the magnetic reconnection occurring above, and creating strong outflows. 

 Figure 3e shows an example of the \ion{Fe}{24} line profile of the flare loop (circle mark in Figure 3a-c) with the Gaussian fit overlaid. The line profile is symmetric and its center is located at almost 192.03 \AA~(stationary) indicating no significant Doppler velocity. Figure 3d shows an example of the \ion{Fe}{24} line profile in the region where the blue-wing is enhanced (+ mark in Figure 3a-c). The profile is a strongly distorted line profile. The distortion extends beyond -400 km s$^{-1}$, which indicates that there are fast plasma flows toward Hinode of 400 km s$^{-1}$. In addition, the width of the line profile is wide which is believed to be a signature of turbulence. Figure 3f shows an example of the \ion{Fe}{24} line profile in the region where the red-wing is enhanced (x mark in Figure 3a-c). This line profile also shows a strongly distorted line profile. The distortion extends beyond 400 km s$^{-1}$, which indicates that there are plasma flows away from Hinode faster than 400 km s$^{-1}$. 

\section{Discussion and Summary}
EIS successfully reveals hot fast flows through the observation of Doppler shifts. These appear as a diffuse structure in the Doppler images ~20 arcsec above the flare loop. At the same time and location we also observe downward flowing loops (apparent speed $\sim$350 km s$^{-1}$). Because the Doppler (EIS LOS direction) and apparent (seen in the AIA images) velocities are similar, the downward flowing loops could be inclined toward the northwest direction ($\sim$45$^{\circ}$). Therefore, we can interpret that the diffuse structure is associated with $\sim$600 km s$^{-1}$ flow perpendicular to the magnetic field. Figure 4 shows the schematic illustration of the observation. The large arrows represent the flow signature observed by EIS Doppler velocity. The small arrows are flows which cannot be observed by EIS Doppler velocity because of LOS effect. The entire flare arcade structure is defined by the combination of Hinode/SDO and STEREO observation (not shown here), the latter of which has a different viewing angle of $\sim$110$^{\circ}$ from the Sun-Earth line. Unfortunately, the STEREO observations were saturated during the main phase of the flare, although the entire flare structure can be determined from the post flare loops. The observed hot fast flow $\sim$20 arcsec above the flare loop is likely to be associated with magnetic reconnection, because it is accompanied with hot fast perpendicular flow. However, the structure of this hot fast flow is diffuse and round-shaped rather than spear-like. This might indicate that the observed hot flows are not the reconnection outflow itself but is illustrating the interaction of the flow with the preexisting flare loop forming a fast-mode shock. Some numerical simulations \citep[e.g.,][]{yok2} show the formation of fast-mode shock above the flare loops. We can speculate that our finding is the hot fast flow in the downstream of fast-mode shock which is caused by reconnection outflow. The signature of the turbulence in the hot fast flow also supports our interpretation. 
\cite{liu} have reported the relationship between the flare outflow and the hard X-ray source. According to their result, the high energy electron (25-50 keV) seems to be located just above the flare loop where the fast flows are terminated. This result may also support the idea that the fast-mode shock could be located just above the flare loop. 
Unfortunately, we do not have hard X-ray observation during our flare event. The relationship between the high energy particles and the diffuse structure which we observe is a topic of future work.
Recently, \cite{ima3} claimed that the outflow region of magnetic reconnection cannot reach ionization equilibrium because of its short Alfv\'en timescale. Thus, in some cases, the reconnection outflow region might be much fainter than we expected. Downstream of the fast-mode shock ionization proceeds much faster because of the high density and temperature conditions. This might be one of the reasons why we can observe the downstream region only. Therefore, we think that the reconnection region is located far above the flare loops where there are very few photons.

\acknowledgments Hinode is a Japanese mission developed and launched by ISAS/JAXA, with NAOJ as a domestic partner and NASA and STFC (UK) as international partners. It is operated by these agencies in co-operation with ESA and NSC (Norway). The Solar Dynamics Observatory is part of NASAfs Living with a Star programme. This work was partially supported by the Grant-in-Aid for Young Scientist B (24740130), by the Grant-in-Aid for Scientific Research B (23340045), by the JSPS Core-to-Core Program (22001).

\begin{figure}
\epsscale{0.7}
\plotone{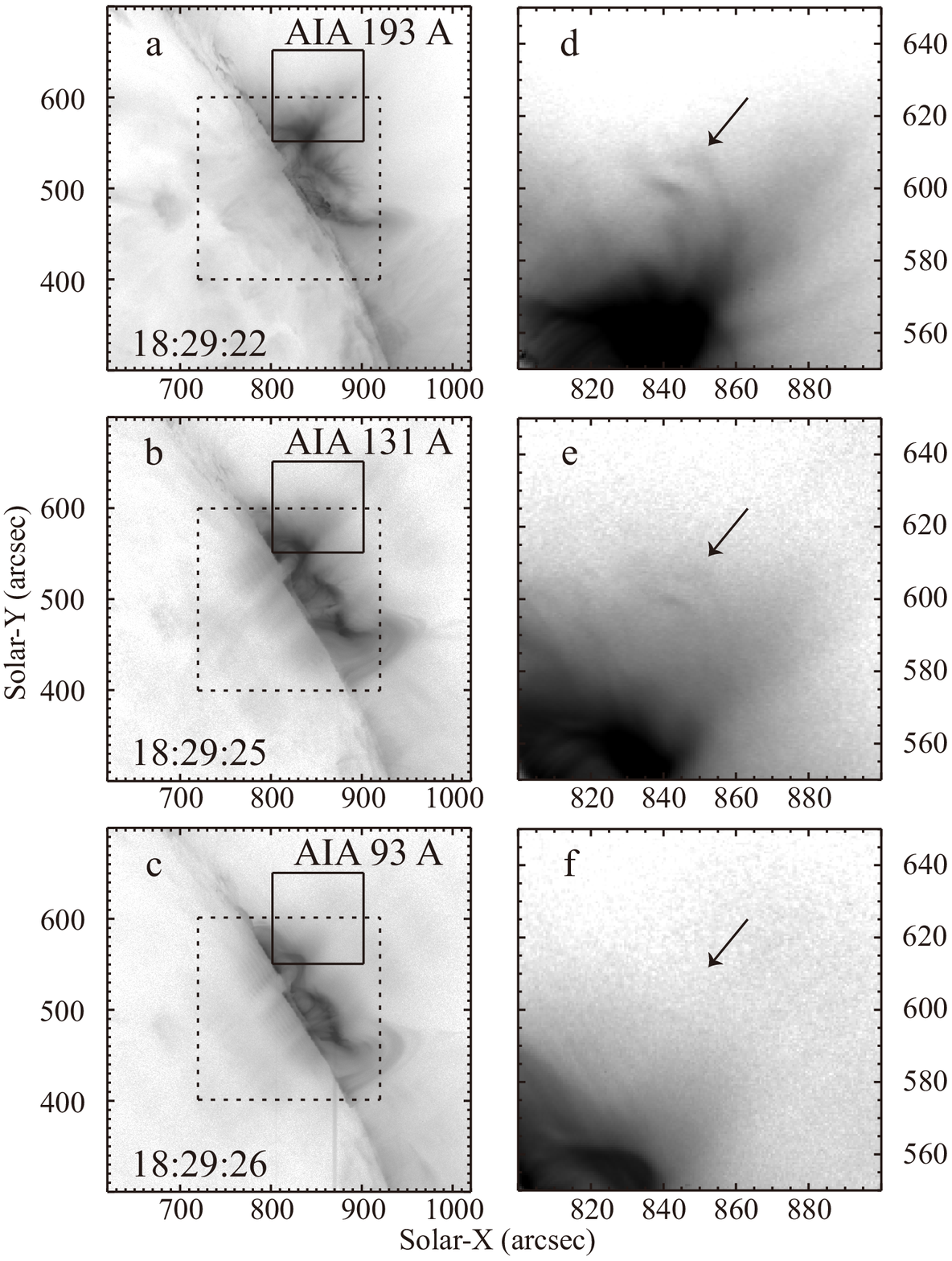}
\caption{Snapshots of the flare on 27 January 2012. The EIS field of view is indicated by the dashed box. (a) AIA 193-\AA~channel image ($10^{7.3}$ K). (b) AIA 131-\AA~channel image ($10^{7.0}$ K). (c) AIA 94-\AA~channel image ($10^{6.8}$ K). (d-f) Enlarged display inside the solid box in a-c. The arrows show the location of downward moving loop.}
\end{figure}

\begin{figure}
\epsscale{0.9}
\plotone{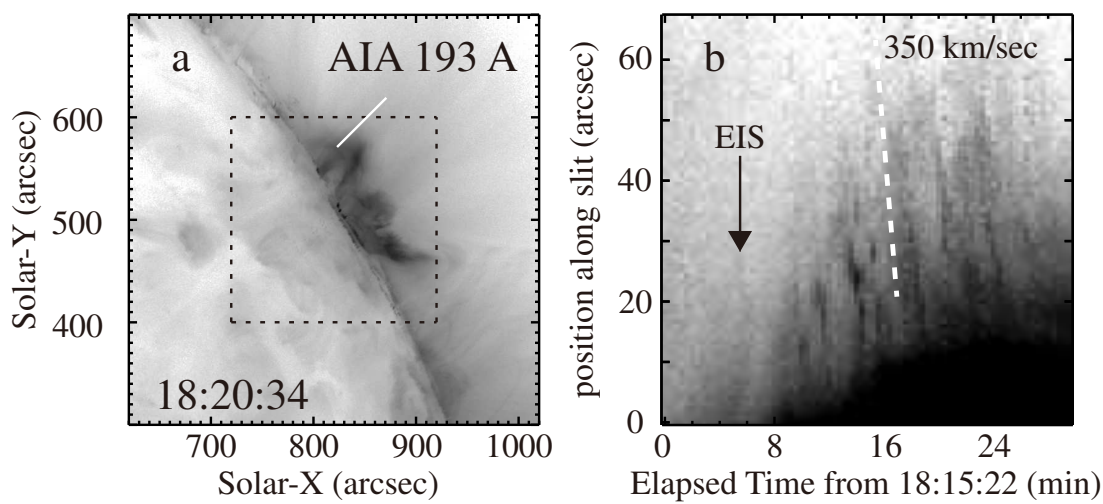}
\caption{(a) AIA 193-\AA~channel image taken at 18:20:34 UT on 27 January 2012 with the position of the CCD pixels (slit) used for preparing (b) shown superimposed as the white bar. The dashed box indicates the EIS filed of view. (b) Time-distance diagram generated from the intensity distribution along the slit shown in (a). The vertical axis shows distance along the slit, and the horizontal axis indicates time in minutes from 27 January 2012, 18:15:22 UT. The dashed line represents transverse velocity of 350 km s$^{-1}$. The arrow shows the observing time of the distorted line profile in Fig. 3d by EIS.}
\end{figure}

\begin{figure}
\epsscale{0.8}
\plotone{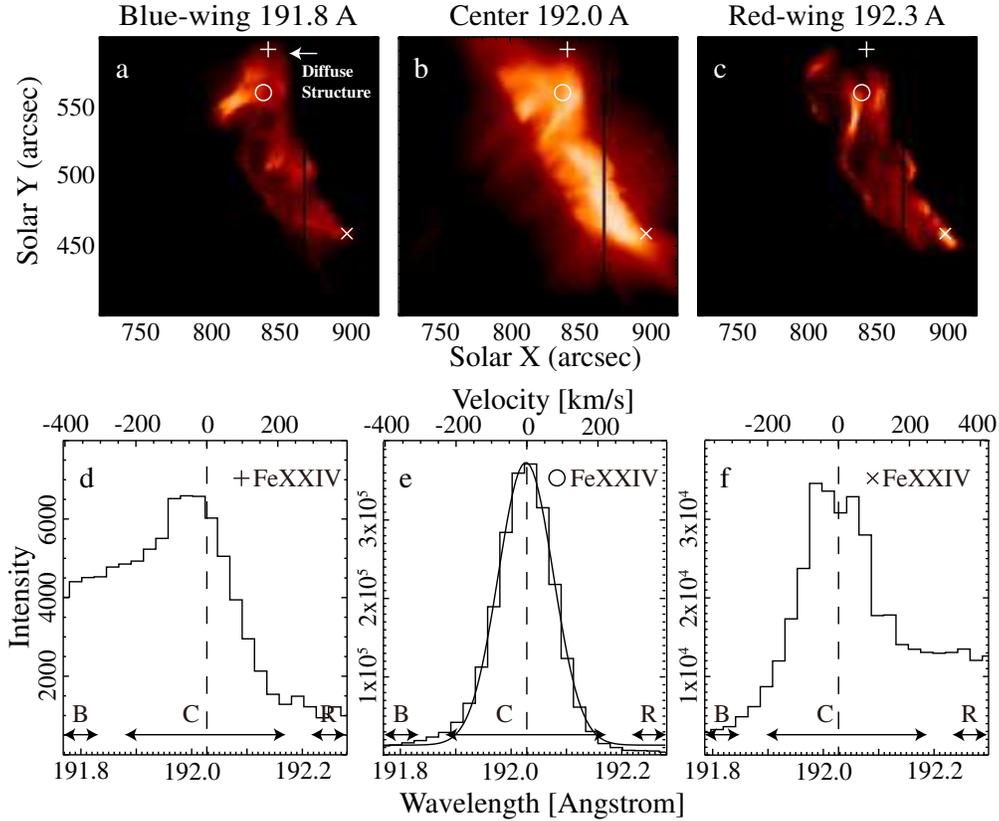}
\caption{(a) EIS \ion{Fe}{24} blue-wing image (191.8 \AA, -400 km s$^{-1}$) of the flare on 27 January 2012. (b) EIS \ion{Fe}{24} center image (192.0 \AA, 0 km s$^{-1}$) of the flare. (c) EIS \ion{Fe}{24} red-wing image (192.3 \AA, 400 km s$^{-1}$) of the flare. (d) The example of \ion{Fe}{24} line profile in the blue-wing enhanced region (+ mark in Fig. 3a-c). The observed timings are marked by arrow in Fig. 2b. (e) The example of \ion{Fe}{24} line profile in the center of the flare (circle mark in Fig. 3a-c). (f) The example of \ion{Fe}{24} line profile in the red-wing enhanced region (x mark in Fig. 3a-c). The ranges B, C, and R in Fig. 3d-f represent the blue-wing, center, and red wing of \ion{Fe}{24}, respectively.}
\end{figure}

\begin{figure}
\epsscale{0.8}
\plotone{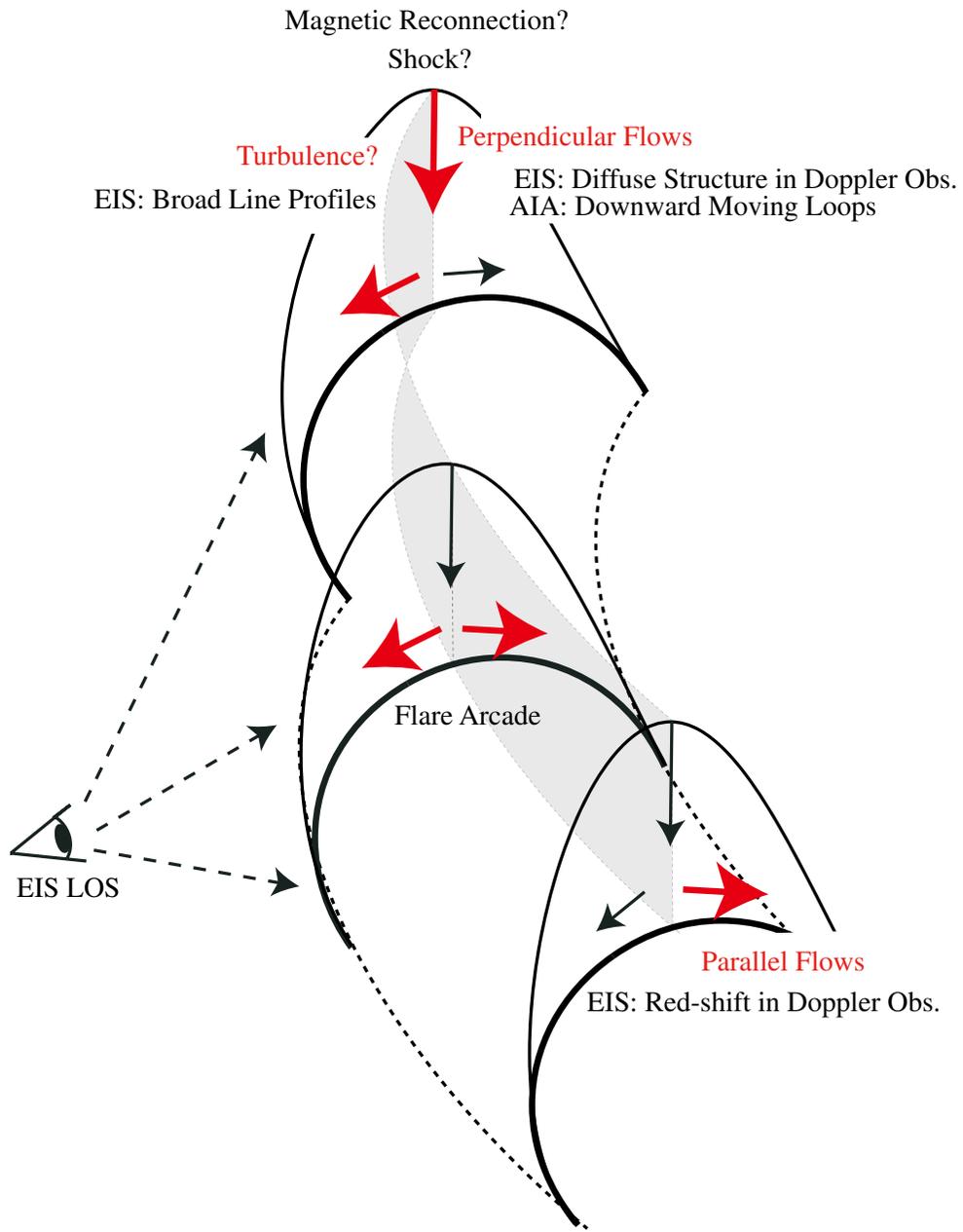}
\caption{ Schematic illustration of the X-class flare on 27 January 2012. The large arrows demark the flow signatures observed by EIS Doppler velocity. The small arrows are flows which cannot be observed by EIS Doppler velocity because of LOS effect. The entire flare arcade structure is defined by the combination of AIA and other view angle observation (STEREO). The figure is illustrated from STEREO viewpoint. EIS and AIA view from left-hand side of the figure.}
\end{figure}

\end{document}